\documentclass[journal]{IEEEtran}
\usepackage{cite}
\usepackage{graphicx}
\usepackage{psfrag}
\usepackage{stfloats}
\usepackage{amsmath}
\usepackage{amssymb}
\usepackage{array}
\usepackage{epsfig}
\long\def\symbolfootnote[#1]#2{\begingroup%
\def\thefootnote{\fnsymbol{footnote}}\footnote[#1]{#2}\endgroup}

\begin{document}
\title{Comments on ``A New Parity-Check Stopping Criterion for Turbo Decoding"}
\author{Yuejun~Wei, Yuhang~Yang, Lili Wei,~\IEEEmembership{Member,~IEEE}, and Wen~Chen,~\IEEEmembership{Senior Member,~IEEE}
\thanks{Manuscript received March 16, 2012. The associate editor coordinating
the review of this letter and approving it for publication was A.~Burr.}
\thanks{The authors are with the Department of Electronic Engineering,
Shanghai Jiao Tong University, Shanghai, 200240, China (e-mail: \{yjwei;~yhyang;~liliwei;~wenchen\}@sjtu.edu.cn).}
\thanks{This work is supported by the National 973 Project 2012CB316106,
2009CB824904, and by NSF China 60972031 and 61161130529.
}} \maketitle
\begin{abstract}
A parity-check stopping (PCS) criterion for turbo decoding is
proposed in \cite{Zhanji:08}, which shows its priority compared with the
stopping criteria of  Sign Change Ratio (SCR), Sign Difference Ratio
(SDR), Cross Entropy (CE) and improved CE-based (Yu) method. But
another well-known simple stopping criterion named
Hard-Decision-Aided (HDA) criterion has not been compared in
\cite{Zhanji:08}. In this letter, through analysis we show that using
max-log-MAP algorithm, PCS is equivalent to HDA; while
simulations demonstrate that using log-MAP algorithm, PCS has
nearly the same performance as HDA.
\end{abstract}
\section{Introduction}
A parity-check stopping (PCS) scheme for iterative turbo decoding is
proposed in \cite{Zhanji:08}, where each soft-input and soft-output
(SISO) decoder outputs both the estimated systematic bits and parity
bits. The systematic bits from one SISO decoder are interleaved and
encoded with the constituent encoder, then the parity bits from the
encoder are compared with the parity bits from another SISO decoder.
If the two sets of parity bits are matched bit by bit, the iterative decoding stops. From the simulation results, the PCS has a smaller average number of iterations compared with the stopping criteria of sign
change ratio (SCR), sign difference ratio (SDR), cross entropy (CE) and an improved CE-based (Yu) methods.

Another well-known stopping criterion named hard-decision-aided
(HDA) has been proposed in \cite{Shao:98}, which compares decisions from the same SISO on successive full-iterations. An improved HDA (IHDA) was proposed in \cite{Nga:01} which requires less storage than HDA with similar performance in terms of bit error rate (BER) and average number of iterations. A general HDA approach was first introduced in \cite{Gracie:99}. This HDA approach compares hard decisions of the systematic bits between SISO1 and SISO2, and can stop the iterative decoding after either SISO, which is currently the best HDA approach. Analysis and simulation results in \cite{Gracie:04} show that the general HDA has very good performance with enhanced max-log-MAP algorithm (i.e., with scaled extrinsic information feedback).

However, the PCS criterion in \cite{Zhanji:08} has not been
compared with the HDA criterion. In this paper, we analyze and compare between PCS and the general HDA criteria, under both max-log-MAP
and log-MAP algorithms. We show that, using max-log-MAP algorithm,  the PCS criterion is equivalent to the
HDA criterion; while using log-MAP algorithm, simulations
demonstrate that both criteria have almost the same performance in terms
of average number of iterations and block error rate (BLER).
\begin{figure*}[!t]
\centerline{\psfig{file=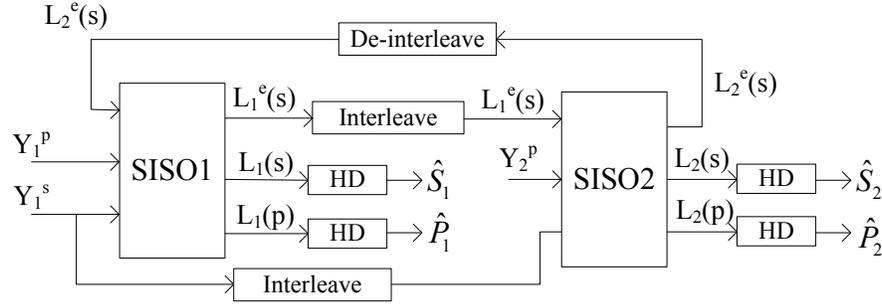,width=4.8in}}
\centering
\caption{Turbo decoding of the systematic bits and the parity bits.}
\hrulefill
\end{figure*}
\section{Comparison Between PCS and HDA Criteria}
In the PCS scheme, each SISO decoder outputs log-likelihood
ratio (LLR) (as shown in Fig.~1) of the systematic bits $s_k$ and
the parity bits $p_k$, which can be expressed as follows,
\begin{eqnarray}\label{for1}
{L}(s_k) &=& \ln\left(\sum_{s',s,
s_k=1}{\alpha_{k-1}(s')\gamma_k(s',s)\beta_k(s)}\right)
\nonumber \\
&&-\ln\left(\sum_{s',s,
s_k=-1}{\alpha_{k-1}(s')\gamma_k(s',s)\beta_k(s)}\right),\nonumber\\
{L}(p_k) &=& \ln\left(\sum_{s',s,
p_k=1}{\alpha_{k-1}(s')\gamma_k(s',s)\beta_k(s)}\right)
\nonumber \\
&&-\ln\left(\sum_{s',s,
p_k=-1}{\alpha_{k-1}(s')\gamma_k(s',s)\beta_k(s)}\right),
\end{eqnarray}
where $\alpha$, $\beta$ and $\gamma$ denote the forward recursive,
backward recursive and branch transition probabilities,
respectively. For max-log-MAP decoding, we can rewrite (\ref{for1})
as:
\begin{eqnarray}\label{eqn02}
{L}(s_k) &=& \max_{s',s, s_k=1}{A_{k-1}(s')\Gamma_k(s',s)B_k(s)}
\nonumber \\
&&-\max_{s',s, s_k=-1}{A_{k-1}(s')\Gamma_k(s',s)B_k(s)},
\nonumber \\
{L}(p_k) &=& \max_{s',s, p_k=1}{A_{k-1}(s')\Gamma_k(s',s)B_k(s)}
\nonumber \\
&&-\max_{s',s, p_k=-1}{A_{k-1}(s')\Gamma_k(s',s)B_k(s)},
\end{eqnarray}
where $A$, $B$ and $\Gamma$ are logarithms of $\alpha$,
$\beta$ and $\gamma$, respectively. Then the hard decisions (HD) of $s_k$ and
$p_k$ are based on the sign of $L(s_k)$ and $L(p_k)$ respectively:
\begin{eqnarray}
\hat s_k = 1, \; \mbox{if} \,\, L(s_k) > 0;\;  \mbox{else} \,\,  \hat
s_k = 0,
\nonumber \\
\hat p_k = 1, \;  \mbox{if} \,\, L(p_k) > 0;\;   \mbox{else} \,\,  \hat
p_k = 0 \label{eqn03:(s,p)}.
\end{eqnarray}
\subsection{Max-log-MAP}
It has been proved in \cite{Marc:98} that, the max-log-MAP algorithm is equivalent to the soft-output Viterbi algorithm (SOVA), and the
max-log-MAP algorithm makes the same hard decisions as the Viterbi algorithm (assuming no tied path metrics). In the decoding process, the max-log-MAP looks at two
paths per step in the trellis, the best with bit zero and the best with bit one, and outputs the difference of the log-likelihoods of the two paths.

From step to step, one of these paths may change, but one will always be the maximum-likelihood (ML) path. That means, for the
max-log-MAP decoding, the hard decision of the output always comes from the ML path because the ML path always has the largest path
metrics (again, assuming no tied path metrics). From (1) and (2), we can see that, the decoding processes
are quite similar for the systematic bits and parity bits.  The difference is the path set of  bit zero and the path set of one.
However, in max-log-MAP decoding, the output of hard decision bits always come from the ML path, which is valid for both systematic bits
and parity bits. Since there is only one ML path in the trellis (without considering the situation of equal path metrics),  $\hat s_k$ and $\hat p_k$ must come from the same path in the
trellis.  In the PCS scheme, there are two parity-check flags to stop the iteration, as shown in Fig.~2:
\begin{enumerate}
\item[(a)] $\hat S_1$ of the $i$-th iteration is interleaved and encoded with the 2nd recursive systematic convolutional encoder (RSC2),
and then compared with $\hat P_2$. If the two sequences are totally matched, the iteration is terminated.
\item[(b)] $\hat S_2$ of the $i$-th iteration is de-interleaved and encoded with RSC1, and then compared with $\hat P_1$. If the two sequences are totally
matched, the iteration is terminated.
\end{enumerate}

\begin{figure}[!t]
\centerline{\psfig{file=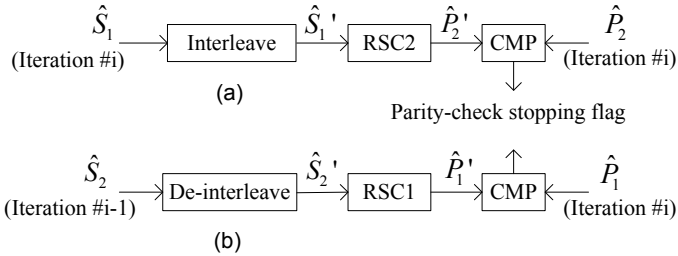,width=3.8in}}
\centering
\caption{The PCS criterion for iterative turbo decoding.}
\end{figure}

\begin{figure}[!h]
\centerline{\psfig{file=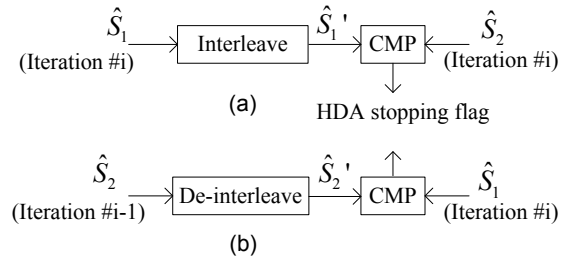,width=3.15in}}
\centering
\caption{The HDA criterion for iterative turbo decoding.}
\end{figure}

The  process of (a) and (b) are similar. For convenience, we take (a) at $i$-th iteration as an example to explain the equivalence
between the PCS and the HDA criteria. As described in the above, for SISO2 decoder, the systematic bits $\hat S_2$  and the parity bits $\hat
P_2$ are in the same ML path in the encoding trellis. So it is easy to deduce that $\hat P_2$  is also the encoded result of $\hat S_2$,
with the encoder of  RSC2. As we know, the RSC encoding is a one-to-one mapping process (assuming a rate 1/2 RSC code), which means
that one information sequence can only give rise to one unique parity sequence, and one parity sequence can only arise from one unique information sequence. So if $\hat P'_2$ is the same as $\hat P_2$, $\hat S'_1$ must be the same as $\hat S_2$. On the other hand, if $\hat S'_1$ is the same as $\hat S_2$, after encoding with RSC2, they must get the same
parity bits. So the PCS criterion is equivalent to comparing $\hat S'_1$ and $\hat S_2$, which is actually the HDA
criterion shown in Fig.~3. Similar to the proof of (a), it's easy to prove that (b) is also equivalent to the HDA criterion.
\subsection{Log-MAP}
Unlike the max-log-MAP and SOVA algorithms which only look at two paths per step, the log-MAP algorithm always takes all paths into
calculation, but splits them into two sets (s=1, s=0), which may change from step to step. The information bits output from the
log-MAP decoder do not certainly constitute a whole path in the trellis. Thus we cannot get a certain relationship between the
systematic bits and parity bits output from a log-MAP decoder. This means that $\hat P_1$ and $\hat P_2$  are not certainly the encoding results of $\hat S_1$ and $\hat S_2$ respectively. However, since log-MAP algorithm and max-log-MAP with a proper scaling factor on the extrinsic information have similar performance \cite{Gracie:99,Vogt:00}, we expect that using log-MAP decoding, the PCS has similar performance with HDA in most scenarios in terms of average iteration number and block error rate. We will perform simulations to compare the two criteria using both max-log-MAP and log-MAP in the next section.

\begin{figure}[t]
\centerline{\psfig{file=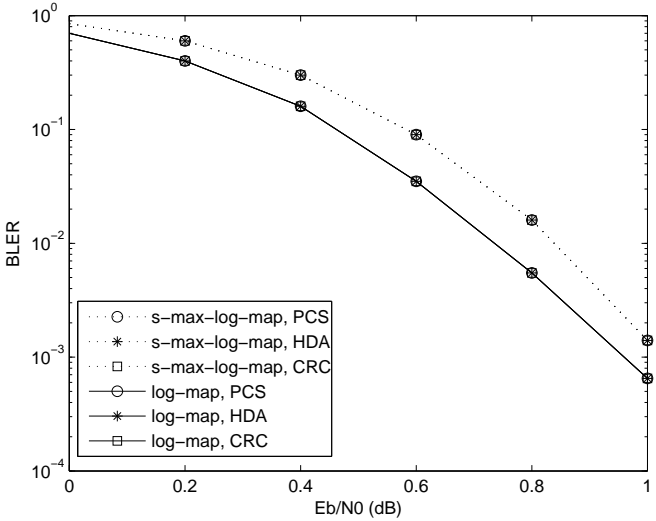,width=3.9in}}
\centering
\caption{The error performance of the log-MAP and max-log-MAP algorithms with different stopping criteria, information length $=990$.}
\end{figure}

\begin{figure}[!t]
\centerline{\psfig{file=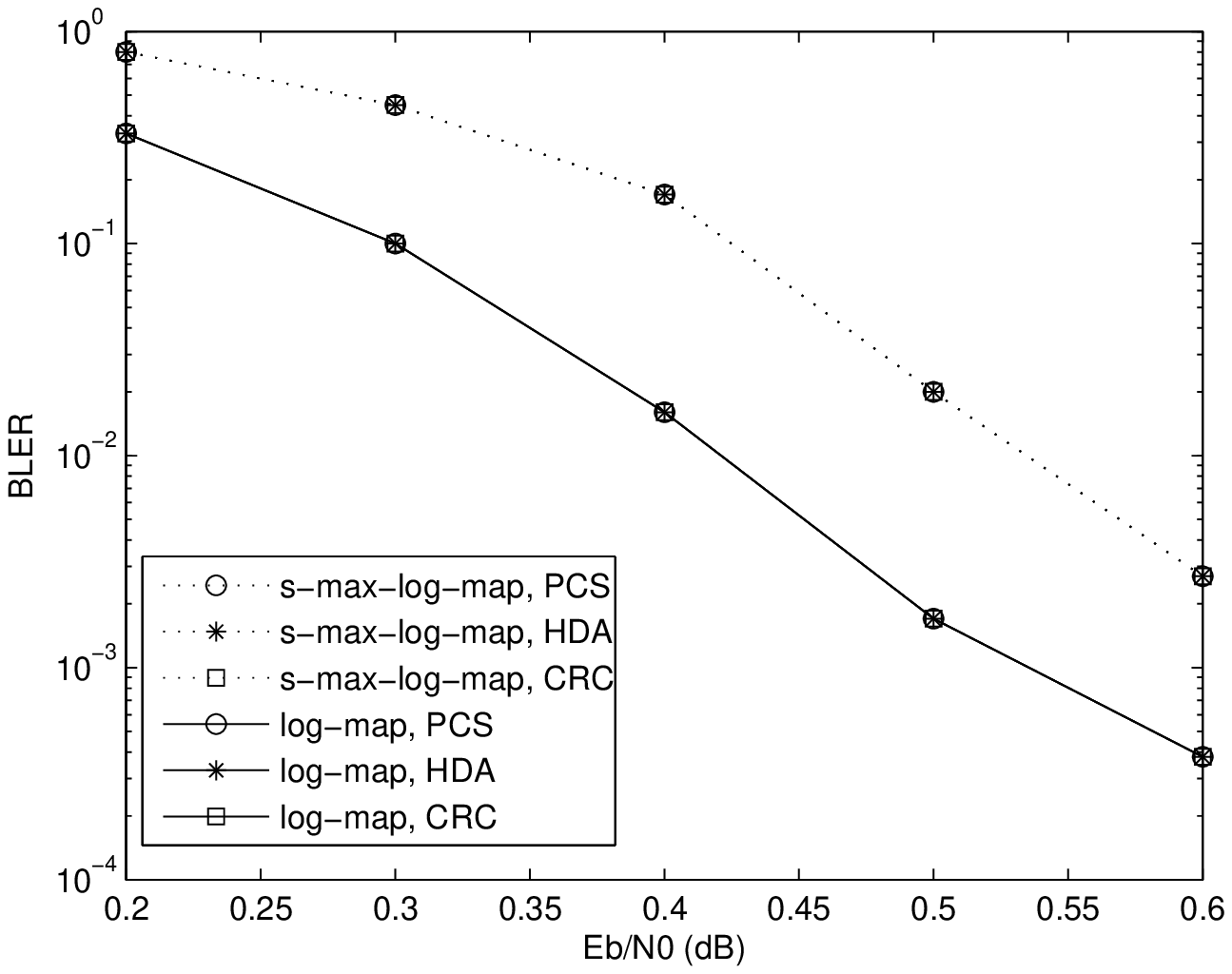,width=3.9in}}
\centering
\caption{The error performance of the log-MAP and max-log-MAP algorithms with different stopping criteria, information length $=5000$.}
\end{figure}

\begin{figure}[!t]
\centerline{\psfig{file=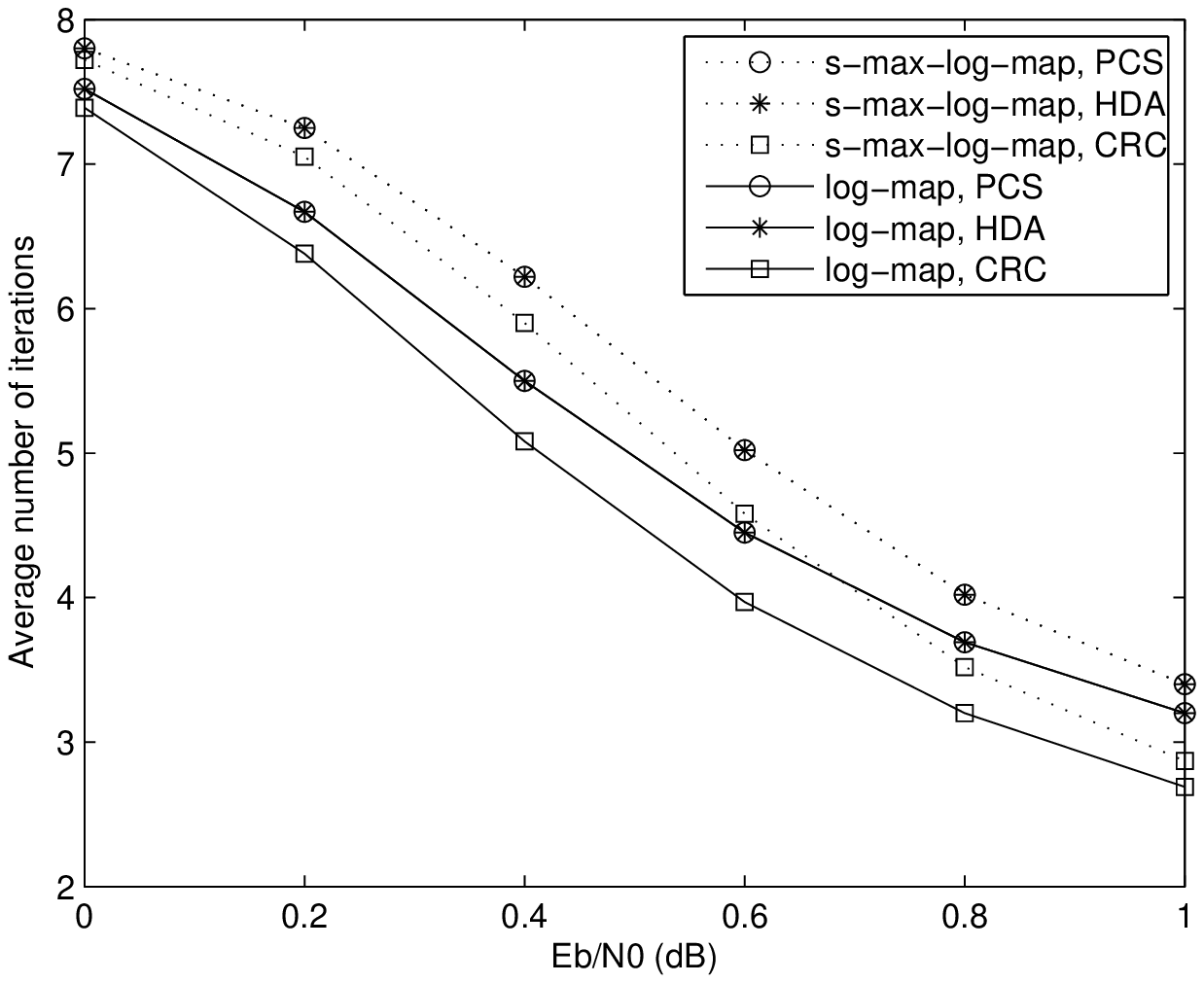,width=3.9in}}
\centering
\caption{Average number of iterations of the log-MAP and max-log-MAP algorithms with different stopping criteria, information length $=990$.}
\end{figure}

\begin{figure}[!h]
\centerline{\psfig{file=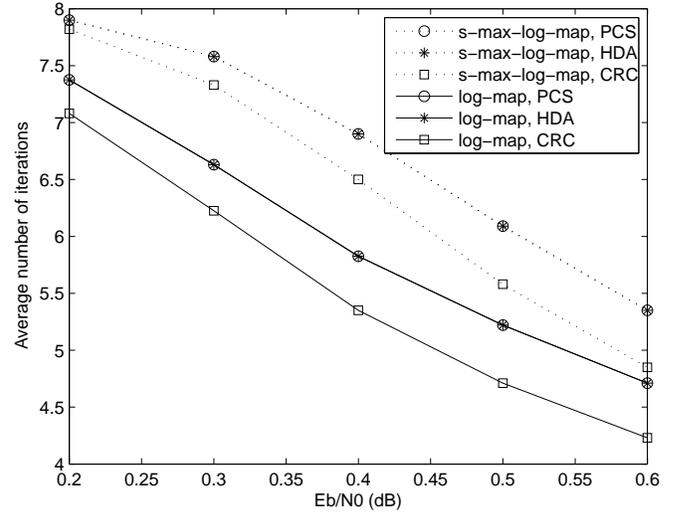,width=3.9in}}
\centering
\caption{Average number of iterations of the log-MAP and max-log-MAP algorithms with different stopping criteria, information length $=5000$.}
\end{figure}

\section{Simulation Results}

Simulation studies are performed with the rate 1/3 turbo code in Universal Mobile Terrestrial Systems (UMTS)~\cite{TS25212:09} in additive white
Gaussian noise (AWGN) channel with binary phase shift keying (BPSK) modulation. The cyclic redundancy check (CRC) stopping scheme with 24-bit CRC is
simulated as the baseline, in which the CRC is checked using the hard decisions of systematic bits after each SISO decoder. The maximum number of decoding
iterations is set to 8 for all simulation cases.

Due to the simplification and approximation in calculating LLRs, the max-log-MAP is suboptimal  and yields an inferior soft output than the log-MAP algorithm. However, the quality of the max-log-MAP algorithm can be improved by using an extrinsic information feedback, referred to as enhanced max-log-MAP in \cite{Gracie:99}, and s-max-log-MAP in \cite{Vogt:00}. With a proper scaling factor, e.g. $0.75$, the performance of s-max-log-MAP algorithm is quite close to log-MAP algorithm. In this simulation, we use the max-log-MAP with an extrinsic scaling factor of $0.75$ and the standard log-MAP algorithms. Of course, the extrinsic information is not scaled for making the decisions.

The path ambiguity (tied path metrics) problem was analyzed in \cite{Gracie:05,Gracie:06}, and it may affect the BER and BLER performance of HDA
early-stopped turbo decoding with finite quantization, especially with coarser quantization. When a soft output is equal to zero, which means that there is
a ``tie" for the best bit decision at that time, then the hard decision on this bit is ambiguous and can be either `1' or `0'. In this letter, we do not permit
early stopping on the current half-iteration if there exists any soft output for hard decision that is exactly zero, which is the same as the solution proposed in \cite{Gracie:05,Gracie:06}.

The simulation results are shown in Fig.~4 - Fig.~7.  From the simulation results, we can see that, for both information length of 990 and 5000, the PCS criterion has the same
performance as HDA criterion in terms of block error rate and average number of iterations for max-log-MAP algorithm.
With log-MAP decoding, two criteria also have nearly the same performance. Comparing with the CRC stopping criterion, both PCS and HDA criteria
have the same block error rate but need more average number of iterations. Since the HDA criterion only need to compare the
information bits, it does not need to check with the parity bits or re-encode the systematic bits, therefore HDA is simpler than PCS.

To make the comparisons above as easy as possible, the 24 bits of overhead required for implementing the CRC stopping rule were just considered part of the information payload. Properly accounting for the reduction in code rate with the 24-bit CRC present requires that the simulation results with CRC stopping be shifted to the right by 10*log10(k/(k-24)) dB, where k is the nominal information block length without the 24-bit CRC. This means that the results for k=990 should be shifted right by about 0.11 dB, and the results for k=5000 should be shifted right by about 0.02 dB. With these shifts included, it can be seen that CRC stopping actually performs worse than the other stopping schemes. However, as shown in \cite{Gracie:04,Gracie:05,Gracie:06}, the performance for the HDA and PCS schemes may start to degrade at lower block error rates with log-MAP decoding.

\section{Conclusions}
In this letter, we analyze and compare the PCS with the HDA stopping criteria. Through analysis we show that using max-log-MAP
decoding, the two criteria are equivalent; using log-MAP decoding,
simulations demonstrate that they have nearly the same performance in
terms of average iteration number and block error rate. Since the
HDA criterion only compares the systematic bits from two constitute
decoders, which does not need to check with the parity bits and
re-encode the systematic bits, it is simpler than the PCS criterion.

\section{Acknowledgement}
The authors would like to thank the anonymous reviewers for their valuable comments and suggestions.


\begin{thebibliography}{1}
\bibitem{Zhanji:08}
Z.~Wu, M.~Peng, and W.~Wang, ``A New Parity-Check Stopping Criterion for
Turbo Decoding," {\em IEEE Commun. Lett.}, vol.~12, no.~4, pp.~304-306, Apr.~2008.
\bibitem{Shao:98}
R.~Y.~Shao, M.~Fossorier, and S.~Lin, ``Two Simple
Stopping Criteria for Iterative Decoding",
{\em Proceedings of the 1998 International Symposium on
Information Theory}, MIT, Cambridge, MA, USA, pp.
279, Aug.~1998.
\bibitem{Nga:01}
T. Ngatched and F. Takawira, ``Simple stopping criterion for turbo
decoding," {\em Electron. Lett.}, vol.~37, pp.~1350-351, Oct.~2001.
\bibitem{Gracie:99}
K.~Gracie, S.~Crozier, and A.~Hunt, ``Performance of a Low-Complexity Turbo Decoder with a Simple
Early Stopping Criterion Implemented on a SHARC Processor,"
{\em Proceedings of the 6th International Mobile Satellite Conference (IMSC '99)},
Ottawa, Ontario, Canada, pp.~281-286, Jun.~1999.
\bibitem{Gracie:04}
K. Gracie, S. Crozier, and P. Guinand, ``Performance of an MLSE-based Early Stopping Technique
for Turbo Codes", {\em Proceedings of the 60th IEEE Vehicular Technology Conference 2004 - Fall (VTC 2004 - Fall)},
Los Angeles, California, USA, Sept.~2004.
\bibitem{Marc:98}
M.~P.~C.~Fossorier, F.~Burkert, S.~Lin and J.~Hagenauer, ``On the
Equivalence Between SOVA and Max-Log-MAP Decodings," {\em IEEE
Commun. Lett.}, vol.~2, no.~5, May~1998.
\bibitem{Vogt:00}
J.~Vogt and A.~Finger, ``Improving the max-log-MAP turbo decoder,"
{\em Electron. Lett.}, vol.~36, no.~23, pp.~1937-1939, Aug.~2000.
\bibitem{TS25212:09}
3GPP TS 25.212 v8.5.0, ``Channel coding and multiplexing," Mar.~2009.
\bibitem{Gracie:05}
K. Gracie, A. Hunt, S. Crozier and P. Guinand,
``MLSE-Based Early Stopping for Turbo Codes - Finite Quantization Effects",
{\em Proceedings of the 2005 Canadian Workshop on Information Theory (CWIT 2005)},
Montr¨¦al, Quebec, Canada, Jun.~2005.
\bibitem{Gracie:06}
K. Gracie, A. Hunt, and S. Crozier, ``Performance of Turbo Codes using MLSE-Based Early
Stopping and Path Ambiguity Checking for Inputs Quantized to 4 Bits",
{\em 4th International Symposium on Turbo Codes \& Related Topics}, Munich, Germany, Apr.~2006.

\end{thebibliography}
\end{document}